\documentclass{article}

\usepackage{spconf}
\usepackage{times}
\usepackage{latexsym}

\usepackage{times}
\usepackage{latexsym}
\usepackage{CJKutf8}

\usepackage[utf8]{inputenc}
\usepackage{graphicx}

\usepackage{inconsolata}

\usepackage[T1]{fontenc}

\usepackage[utf8]{inputenc}

\usepackage{microtype}

\usepackage{math-common}
\usepackage{makecell}

\usepackage{xcolor}
\usepackage{inconsolata}
\definecolor{chromeyellow}{rgb}{1.0, 0.65, 0.0}

\newcommand{\nop}[1]{}

\usepackage[T1]{fontenc}

\usepackage[utf8]{inputenc}

\usepackage{microtype}

\usepackage{inconsolata}

\usepackage{colortbl,booktabs}
\definecolor{mygray}{gray}{0.95}

\usepackage{amssymb}
\usepackage{pifont}

\usepackage{numprint,siunitx,xcolor}
\usepackage{multirow}
\usepackage{tabu}
\usepackage{graphics}
\usepackage{float}
\usepackage{lipsum}

\renewcommand{\paragraph}[1]{\noindent\textbf{#1}}

\newcommand\blfootnote[1]{%
  \begingroup
  \renewcommand\thefootnote{}\footnote{#1}%
  \addtocounter{footnote}{-1}%
  \endgroup
}

\usepackage{xparse}

\title{In-Context Prompt Editing for Conditional Audio Generation}
\name{Ernie Chang$^{*^\spadesuit}$, Pin-Jie Lin$^{*\clubsuit}$, Yang Li$^{\spadesuit}$, Sidd Srinivasan$^{\spadesuit}$, Gael Le Lan$^{\spadesuit}$,}
\myname{David Kant$^{\spadesuit}$, Yangyang Shi$^{\spadesuit}$, Forrest Iandola$^{\spadesuit}$, Vikas Chandra$^{\spadesuit}$}

\address{$^\spadesuit$Meta AI\\
         $^\clubsuit$Language Science and Technology, Saarland University\\
         \texttt{\{erniecyc, yangli1, siddsrinivasan, davidkant\}@meta.com} \\
         \texttt{pinjie@lst.uni-saarland.de}}
\begin{document}
%
\maketitle

\begin{abstract}

\emph{Distributional shift} is a central challenge in the deployment of machine learning models as they can be ill-equipped for real-world data.
This is particularly evident in text-to-audio generation where the encoded representations are easily undermined by unseen prompts, which leads to the degradation of generated audio --- the limited set of the text-audio pairs remains inadequate for conditional audio generation in the wild as user prompts are under-specified.
In particular, we observe a consistent audio quality degradation in generated audio samples with user prompts, as opposed to training set prompts.
To this end, we present a \emph{retrieval-based in-context prompt editing} framework that leverages the training captions as demonstrative exemplars to revisit the user prompts.
We show that the framework enhanced the audio quality across the set of collected user prompts, which were edited with reference to the training captions as exemplars.

\end{abstract}
\newtheorem{cor}{Corollary}
\newtheorem{condition}{Condition}
\newcommand{\tmix}{t_{\text{mix}}}
\newcommand{\ar}[1]{\textcolor{blue}{AR:{#1}}}

\newcommand{\IS}{DSIR\xspace}
\newcommand{\qfeat}{q_{\text{feat}}}
\newcommand{\pfeat}{p_{\text{feat}}}
\newcommand{\hatqfeat}{\hat{q}_{\text{user}}}
\newcommand{\hatpfeat}{\hat{p}_{\text{train}}}
\newcommand{\primepfeat}{{p'}_{\text{new}}}

\newcommand{\lambdamax}{\lambda_{\text{max}}}
\newcommand{\onehot}{\text{one-hot}}
\newcommand{\lambdamin}{\lambda_{\text{min}}}
\newcommand{\fisherinfj}{I_{j,\thetastar}}
\newcommand{\fisherinfk}{I_{k,\thetastar}}
\newcommand{\conditionnum}{\gamma_{\thetastar}}
\newcommand{\fn}{f_n}
\newcommand{\rnthetastar}{r_n(\thetastar)}
\newcommand{\rntheta}{r_n(\theta)}
\newcommand{\dsname}{GINC\xspace}
\newcommand{\Topic}{Concept\xspace}
\newcommand{\Topics}{Concepts\xspace}
\newcommand{\topic}{concept\xspace}
\newcommand{\topics}{concepts\xspace}
\newcommand{\hiddenseg}{H}
\newcommand{\hiddensegstart}{h^{\text{start}}}
\newcommand{\hiddensegstarttest}{h^{\text{start}}_{\text{test}}}
\newcommand{\hiddensegstarttestprime}{{{h^{\text{start}}_{\text{test}}}'}}
\newcommand{\hiddensegend}{H^{\text{seg}}}
\newcommand{\hiddensegendtest}{H^{\text{seg}}_{\text{test}}}
\newcommand{\promptseq}{S_n}
\newcommand{\obsex}{O^{\text{ex}}}
\newcommand{\Lzeroone}{L_{\text{0-1}}}
\newcommand{\LCE}{L_{\text{CE}}}
\newcommand{\indicator}{\mathbf{1}}
\newcommand{\badset}{\sB}
\newcommand{\startconst}{\frac{1}{|\sD|}}
\newcommand{\delimub}{c_2}
\newcommand{\delimlb}{c_1}
\newcommand{\delimstartub}{c_4}
\newcommand{\delimstartlb}{c_3}
\newcommand{\translb}{c_5}
\newcommand{\emitlb}{c_6}
\newcommand{\examplelb}{c_7}
\newcommand{\hiddenstartlb}{c_8}

\newcommand{\KLk}{KL_k(\thetastar \| \theta)}
\newcommand{\KLj}{KL_j(\thetastar \| \theta)}
\newcommand{\KLjprev}{KL^\theta_{j-1}}
\newcommand{\KLjnext}{KL^\theta_{j+1}}
\newcommand{\ppromptj}{\pprompt^j}
\newcommand{\pjtheta}{p^j_\theta}
\newcommand{\pjnexttheta}{p^{j+1}_\theta}
\newcommand{\pjthetastar}{p^j_\thetastar}
\newcommand{\pjnextthetastar}{p^{j+1}_\thetastar}
\newcommand{\ptwojtheta}{p^{(2,j)}_\theta}
\newcommand{\pijtheta}{p^{(i,j)}_\theta}
\newcommand{\ptwojthetastar}{p^{(2,j)}_\thetastar}
\newcommand{\pijthetastar}{p^{(i,j)}_\thetastar}
\newcommand{\ptwojnexttheta}{p^{(2,j+1)}_\theta}
\newcommand{\pijnexttheta}{p^{(i,j+1)}_\theta}
\newcommand{\ptwojnextthetastar}{p^{(2,j+1)}_\thetastar}
\newcommand{\pijnextthetastar}{p^{(i,j+1)}_\thetastar}

\newcommand{\errstart}{\epsilon^\theta_{\text{start}}}
\newcommand{\errdelim}{\epsilon^\theta_{\text{delim}}}

\newcommand{\h}{h}
\newcommand{\obs}{o}
\newcommand{\obsset}{\sO}
\newcommand{\delim}{h^{\text{delim}}}
\newcommand{\obsseg}{O}
\newcommand{\obsdelim}{o^{\text{delim}}}
\newcommand{\X}{x}
\newcommand{\y}{y}
\newcommand{\Xtest}{x_{\text{test}}}
\newcommand{\ytest}{y_{\text{test}}}
\newcommand{\bigK}{K}

\newcommand{\hatp}{\hat{p}}
\newcommand{\pprompt}{p_{\text{prompt}}}
\newcommand{\ppromptstart}{p_{\text{prompt}}}
\newcommand{\ppromptdelim}{p_{\text{prompt}}^{\text{delim}}}
\newcommand{\unif}{\text{Unif}}
\newcommand{\betax}{\beta(x_{n+1})}

\newcommand{\Tic}{\sT_{\text{in-context}}}

\newcommand{\minv}{{-1}}
\newcommand{\softmax}{\text{softmax}}
\newcommand{\minimize}{\text{minimize}}
\newcommand{\thetastar}{{\theta^*}}
\newcommand{\statdistpiH}{\pi^\thetastar}
\newcommand{\statdistpiHx}{\pi_x^\thetastar}
\newcommand{\statdistpiHtheta}{\pi^\theta}
\newcommand{\statdistpiHthetax}{\pi_x^\theta}
\newcommand{\statdistpiD}{\pi^\thetastar_\sD}
\newcommand{\statdistpiDtheta}{\pi^\theta_\sD}
\newcommand{\conetheta}{C_1^\theta(\epsilon^\theta(t_n))}
\newcommand{\ctwotheta}{C_2^\theta(\epsilon^\theta(t_n))}
\newcommand{\conethetastar}{C_1^\thetastar(\epsilon^\thetastar(t_n))}
\newcommand{\ctwothetastar}{C_2^\thetastar(\epsilon^\thetastar(t_n))}
\newcommand{\conethetat}{C_1^\theta(\epsilon^\theta(t))}
\newcommand{\ctwothetat}{C_2^\theta(\epsilon^\theta(t))}
\newcommand{\conethetastart}{C_1^\thetastar(\epsilon^\thetastar(t))}
\newcommand{\ctwothetastart}{C_2^\thetastar(\epsilon^\thetastar(t))}

\newcommand{\tilf}{\tilde{f}}
\newcommand{\maximize}{\text{maximize}}
\newcommand{\spoc}{\textsc{SPoC}}
\newcommand{\testp}{\textsc{TestP}}
\newcommand{\testw}{\textsc{TestW}}
\newcommand{\Pibeta}{\Pi_{\beta}}
\newcommand{\hatalpha}{\hat{\alpha}}
\newcommand{\fstd}{f_{\text{direct}}}
\newcommand{\fstdi}{f_{\text{direct},j}}
\newcommand{\fhatpi}{\hat{f}_{\Pi}}
\newcommand{\fpi}{f_{\Pi}}
\newcommand{\hattheta}{\hat{\theta}}
\newcommand{\ftheta}{f_{\theta}}
\newcommand{\fthetai}{f_{\theta_j}}
\newcommand{\fthetamin}{f_{\theta_{\text{min}}}}
\newcommand{\fthetabasemin}{f^*_{\text{base}}}
\newcommand{\fthetabase}{f_{\text{base}}}
\newcommand{\fthetacomp}{f_{\text{comp}}}
\newcommand{\fhattheta}{\hat{f}_{\theta}}
\newcommand{\fhat}{\hat{f}}
\newcommand{\fstar}{f^\star}
\newcommand{\proj}{\text{proj}}
\newcommand{\projC}{\text{proj}_{C}}
\newcommand{\sign}{\text{sign}}
\newcommand{\thetareg}{{\theta_{\text{reg}}}}
\newcommand\uy{\tilde y}
\newcommand\noiseuy{\tilde z}

\newcommand{\yone}{\y_{\text{max}}}

\newlength{\widebarargwidth}
\newlength{\widebarargheight}
\newlength{\widebarargdepth}
\DeclareRobustCommand{\widebar}[1]{%
    \settowidth{\widebarargwidth}{\ensuremath{#1}}%
    \settoheight{\widebarargheight}{\ensuremath{#1}}%
    \settodepth{\widebarargdepth}{\ensuremath{#1}}%
    \addtolength{\widebarargwidth}{-0.3\widebarargheight}%
    \addtolength{\widebarargwidth}{-0.3\widebarargdepth}%
    \makebox[0pt][l]{\hspace{0.3\widebarargheight}%
    \hspace{0.3\widebarargdepth}%
    \addtolength{\widebarargheight}{0.3ex}%
    \rule[\widebarargheight]{0.95\widebarargwidth}{0.1ex}}%
{#1}}

\newcommand{\shibani}[1]{\todo[color=blue!40]{SS: #1}}

\blfootnote{$^\ast$ Equal contribution.}

\section{Introduction}

Recently, there has been notable progress in the task of conditional text-to-audio (TTA) generation, where audio signals can be synthesized from textual descriptions~\cite{yang2023diffsound, liu2023audioldm}. 
In most setups, text encoders model text prompts as priors for audio decoders to condition upon, and rely heavily on the amount of parallel text-audio data for generalizability.
Consequently, TTA models' adaptability is constrained to the training prompt distributions which were accessible during training, and collecting data from all possible prompt distributions is impractical.

Thus, one major limitation remains as the limited ability to generalize across the distribution shift.
This shift in text distribution in the wild diverges from the training captions that the model has been trained on, resulting in an inadequately equipped text encoder and leading to inaccuracies in representing unseen textual inputs. 
These inaccuracies further cascade into errors during the subsequent decoding inference steps, hindering the overall synthesis quality.
Empirically, we observe a marked audio quality degradation (see Figure~\ref{fig:dist}) when there is a distributional shift from the training prompt distribution $P_{t}(x)$ to the user prompt distribution $P_{u}(x)$.
The reason is that the learned text representation $P(z|x; \theta)$ remains constant while the acquired prior $\theta$ is unable to adapt to unseen distribution.
Thus, this tendency for models to have better audio quality within the training prompt distribution hinders the ability of the model to be deployed in real-world settings, as it is impossible to train the model on all possible data distributions that it may encounter in the real world.

In this paper, we first discuss the distributional shift in deployed conditional audio generation systems (Section~\ref{prompt_div}).
We observed that the shift in prompts leads to lower audio quality measured in terms of FAD~\cite{kilgour2019frechet} and CLAP~\cite{elizalde2023clap}.
To handle this shift, we propose to edit the user prompts with instruction-tuned large language models (LLMs) (i.e., LLaMA 2~\cite{touvron2023llama}). 
However, using LLMs as-is results in ill-formulated prompting, which can lead to sub-par performance~\cite{min2022rethinking,wei2022chain}. 
The use of \emph{demonstrative exemplars} for large language models~\cite{DBLP:journals/corr/abs-2005-14165,touvron2023llama} has recently been shown to bridge the gap between seen and unseen prompts. 
To this end, we introduce a framework for LLM-based prompt editing with demonstrative exemplars. To validate our approach, we conducted extensive experiments on collected user prompts consisting of a range of free-form entered texts.

We summarize our contributions as follows:
\begin{itemize}
    \item First, we put forward a way to quantify the distributional shift in prompts with feature-based KL divergence reduction. We  compute this distributional ``prompt divergence'' and establish its correlation to the audio quality in terms of FAD scores.
    \item We adopted in-context learning from text-only usages to text-conditioned audio generation, and show that in-context prompt editing enhances the audio quality across a range of evaluation metrics, including CLAP, Fréchet audio distance (FAD), and human evaluation.
    \item We improved upon the  computationally expensive of prompt retrieval from large-scale datasets. This is achieved via de-duplication of the training prompts with the minHash algorithm, then using the K-means clustering technique to split prompts into groups for fast retrieval of  relevant exemplars. 
\end{itemize}



\section{Preliminaries on Prompt Distribution}
\label{prompt_div}
To model two text prompt distributions, we need to first project them into features where additional metrics can be computed. 
Here we first denote the training and user prompt distribution as $\{P_t, P_u\}$ respectively, and formalize the following feature extraction process: 
(1) Given a context (or prompt) $x_u$ from user prompt data $D_u$, we retrieve a prompt $x_{t}$ from training distribution $P_t$, or joint distribution with language models $P_{t} \cup P_{LM}$, which we will elaborate in Section~\ref{sec:approach}. (2) Conditioned on prompt $x_{t}$, we sample the latent representation $z \backsim f(x_{t})$ from trained text encoder $f(\cdot)$. 


Note that $f(\cdot)$ is a generalized text encoder, which can be any pretrained text encoder such as RoBERTa~\cite{liu2019roberta}, T5~\cite{raffel2020exploring}, or CLAP~\cite{wu2023large}. 
As such, the metric is suitable for any text encoder models, which makes the approach rather generalized.

\begin{figure}[h]
  \centering
\includegraphics[width=0.8\columnwidth]{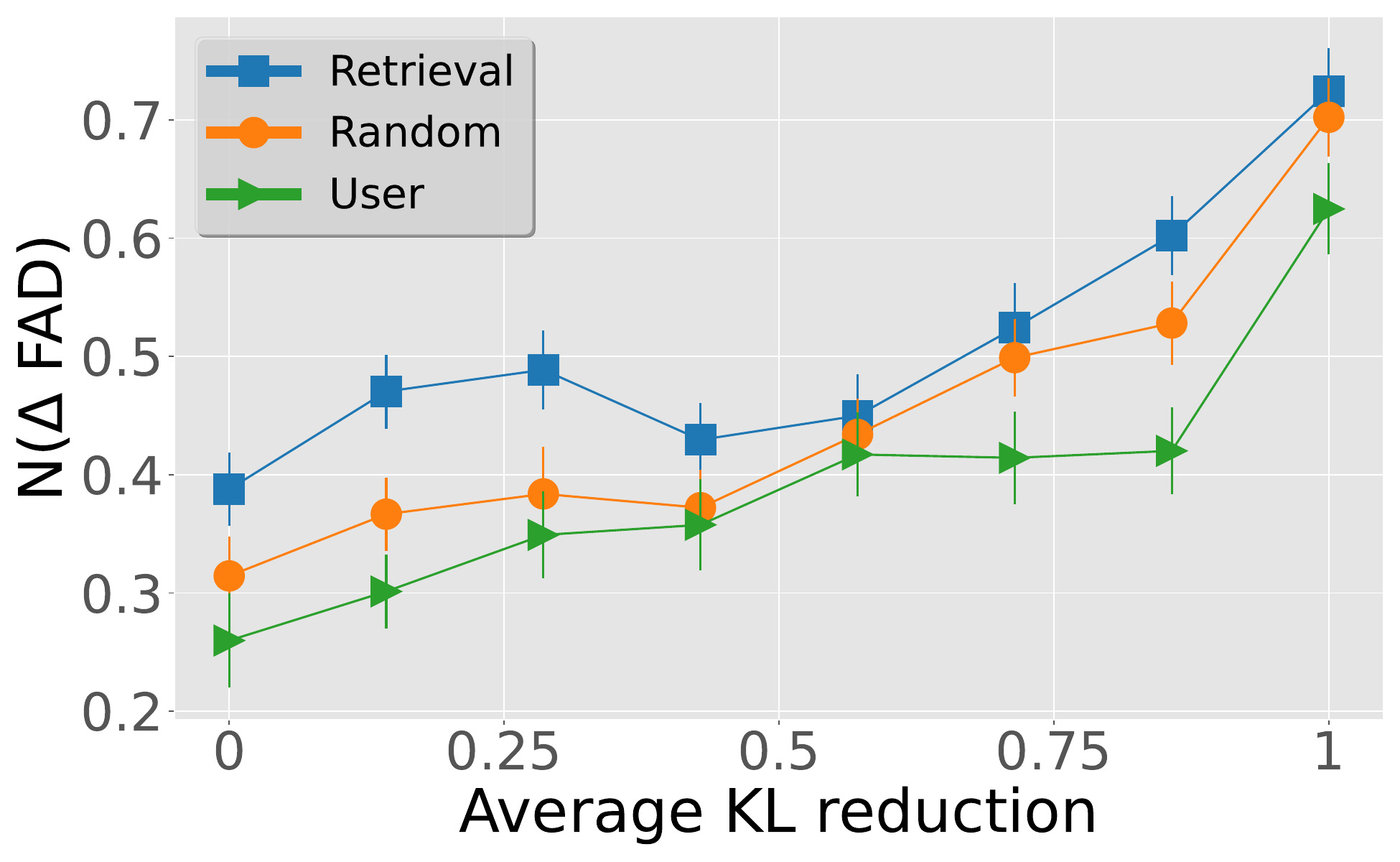}
\caption{ \small 
Plot of average KL reduction on the n-gram feature space, defined as how much the selected dataset reduces KL divergence to the target distribution over just random sampling. 
The \emph{retrieval} uses the data samples from the training prompt distribution, and the \emph{user} specifies the input from the user prompt distribution.
There is a strong correlation between KL reduction and FAD reduction.
}
\label{fig:dist}
\end{figure}

\paragraph{Distributional shift in prompts.} We propose to measure the KL reduction as it reflects the relative divergence when user prompts are fed to the text encoder as opposed to the training set prompts~\cite{xie2023data}. 
Instead of measuring divergence at the text level, we utilize the encoded text's feature distribution $Z$. 
Here we define the Kullback–Leibler divergence ($\text{KL}(P||Q)$) between two encoded text features $P(X)$ and $Q(X)$:
\begin{equation}
\small
    \text{KL}(P||Q) = \frac{1}{|X|}  \sum_{x \in X} P(x) \log \frac{P(x)}{Q(x)}.
\end{equation}
where each input $x$ consists of normalized scores bounded in $[0,1]$. 
The KL value is then averaged across feature channels; and the prompt divergence score $r_{div}$ is given as:

\begin{align}
r_{div} = \frac{1}{|Z|} \sum_{\hatpfeat \in Z, } \text{KL}(\hatpfeat \| \hatqfeat) - \text{KL}(\hatpfeat \| \primepfeat)
\end{align}
where $\hatpfeat$ is the original prompt feature distribution, $\hatqfeat$ is the converted prompt feature distribution, and $\primepfeat$ is the empirical feature distribution of the sampled prompt $x_{t}$.
$Z$ is the set of extracted text features (or latent code $z$)  with text encoder $f(\cdot)$.
We then examine the relationship between the {KL reduction} induced by a specific prompt editing approach to the resulting audio quality in terms of FAD scores in Figure~\ref{fig:dist}. 
A smaller $r_{div}$ value indicates that the prompt-induced audio distribution captures the in-domain audio quality, under retrieved prompt distribution $P_t$.

\begin{figure*}[ht]
\centering
\includegraphics[width=0.95\textwidth,height=1\textheight,keepaspectratio]{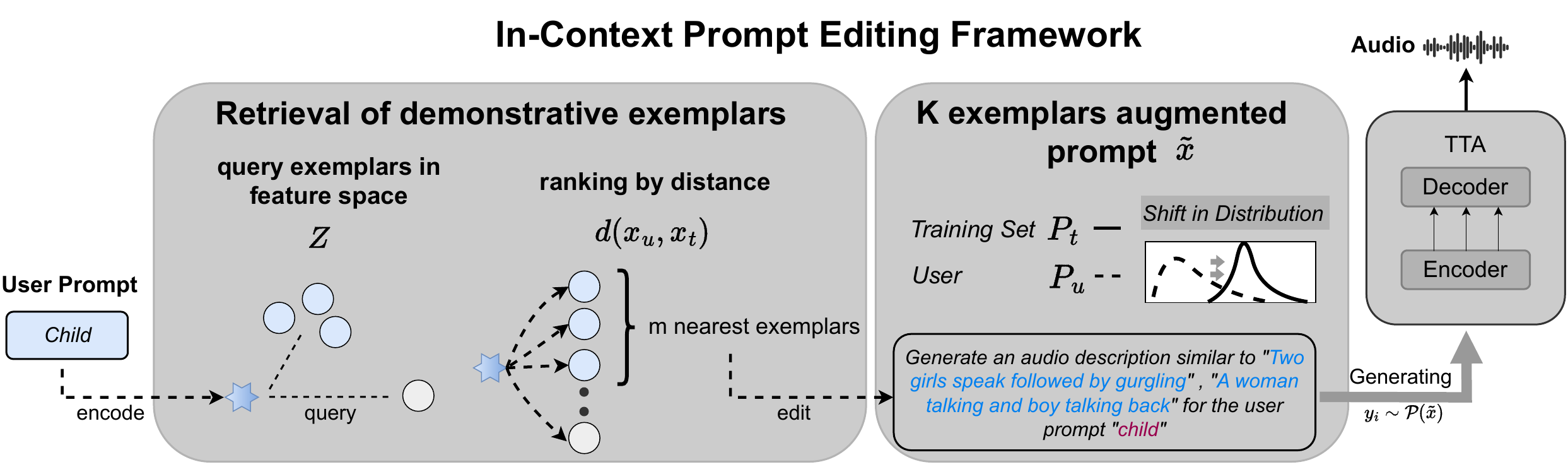}
\caption{ \small Diagram depicting the process of in-context prompt editing for improved audio quality. 
Training set is first clustered via K-means, then top-$M$ prompts are retrieved based on user queries, of which the most similar prompt is then used as the exemplar for in-context prompt editing with LLM. 
Prior to finding representative centroids, we apply de-duplication to eliminate the nearly identical demonstrative examples in the training set. This enable us to retain sufficient data to represent the data distribution while improve retrieval efficiency.
}
\label{fig:main}
\end{figure*}


\section{In-Context Prompt Editing}
\label{sec:approach}
Inspired by the recent successes in in-context learning~\cite{DBLP:journals/corr/abs-2005-14165}, we formulate the prompt editing process as follows:
Let $x_u$ be a query input prompt, written by the user, and consider $Y = \{y_1, \ldots , y_m\}$ as the set of refined audio samples.
We edit $x_u$ by incorporating a task instruction $I$ and an in-context demonstration set $C = \{c_1, \ldots , c_k\}$, which consists of $k$ demonstration examples.
Each $c_i$ is a caption retrieved from training prompt set $D_t$, and the resultant in-context prompt is formulated as $\tilde{x}= [I, C, x_u]$. 
We approximate the likelihood of the audio $y_j$ being representative using a scoring function $f$ parameterized by $\theta$ and applied to the entire input sequence:
 
\begin{equation}
    P(y_j | x_u) \triangleq f_{\theta}(y_j, \tilde{x}; \theta)
\end{equation}

Rather than conditioning on an unseen user prompt, we draw the audio signal from a surrogate distribution: $\hat{y} = \arg \max P_{y_j \in Y}(y_j | \tilde{x}; \theta)$.
Given the challenges posed by distributional shift arising from disparities between training and real distributions, we present a framework for in-context prompt editing.
The framework edit user prompts into with demonstrative exemplars from the training prompt distribution.
Primarily, the process of editing in-context prompts based on a collection of training prompts $D_t$  consists of two major steps:

\begin{enumerate}
    \item \emph{De-duplication} to improve retrieval efficiency, since the data $D_t$ can be prohibitively large.
    \item \emph{Retrieval of demonstrative exemplars} for language model inference.
\end{enumerate}

In what follows, we provide the details of these steps.

\subsection{De-duplication} 
Retrieving prompts from large-scale datasets can result in resource-intensive computations, especially when multiple  pairs of similar documents are present in the data.
Thus, the goal of de-duplication is to eliminate duplicate or nearly identical items from a large sample pool.
To do so, we adopt MinHash \cite{10.5555/829502.830043} for identifying demonstrative exemplars within the training dataset. 
MinHash represents each document, denoted as $x_i$ and $x_j$, using sets of $n$-grams, expressed as $d_i$ and $d_j$ respectively. 
The similarity between these sets is measured using the Jaccard Index \cite{jaccard1912distribution} to indicate the overlap between the sets.
We discard high Jaccard indexes which are highly matched documents for similarity greater than $0.8$.

\subsection{Retrieval of demonstrative exemplars} 
The retrieval process begins with K-means clustering using the Faiss library~\cite{johnson2019billion} which is built for efficient similarity search. 
Each training text prompt is projected into embeddings with the sentence encoder (S\-BERT~\cite{reimers2019sentencebert}). 
For the sake of ease of analysis, we use AudioCAPS~\cite{kim-etal-2019-audiocaps} and BBC sounds~\cite{liu2023audioldm} training prompts as exemplars for in-context prompts in clustering. 

For each user prompt $x_u$, we first perform similarity search with the indexed clusters and obtain the top-$M$ closest neighbors -- $x_1, x_2, ..., x_k$ -- from the training set using the distances within the sentence encoder's embedding space. 
Utilizing these neighboring exemplars, the in-context demonstration set $C$ is constructed, wherein each $x_i$ corresponds to $c_i$.
We order the neighbors to satisfy $d(c_1) \leq d(c_k)$ when $i < j$. 
This ranking provides a natural hierarchy of sentences within the cluster, based on their contextual relevance to the user's query. 
Consequently, the top-$M$ candidate prompts are selected as illustrative examples.
We then structured the top candidate as in Figure~\ref{fig:main}.

\begin{table*}[tbp]
\centering
\small
\begin{tabular}{l|c|ccc|cc}
\toprule
    \textbf{Prompting Approach}          & $\Delta r_{div}$~$\uparrow$ & $\Delta$ CLAP~$\uparrow$  & $\Delta$ KL~$\uparrow$   & $\Delta$ FAD~$\uparrow$ & SUB~$\uparrow$ & OBJ~$\uparrow$   \\
\midrule
User                                   & -     & -      & -    & -   & 3.58 & 1.54 \\
\midrule
Random                                & 0.472 & 0.003 & 0.1013 & 0.590 & 3.65 & 2.56 \\
LLM-only                              & 0.044 & 0.036 & 0.0649 & 0.943 & 3.61 & 2.58 \\
\midrule
exemplar (out-domain, K=100, random)  & 0.444 & 0.019 & 0.0660 & 1.433 & 3.66 & 2.61  \\
exemplar (in-domain, K=100, random)   & 0.439 & 0.025 & 0.0469 & 1.803 & 3.72 & 2.62 \\
exemplar (in-domain, K=100, farthest) & 0.464 & 0.046 & 0.0577 & 2.203 & 3.78 & 2.64 \\
exemplar (in-domain, K=50, closest)   & 0.520 & 0.042 & 0.0813 & 2.860 & \textbf{3.84} & \textbf{2.69} \\
exemplar (in-domain, K=100, closest)  & \textbf{0.594} & \textbf{0.047} & \textbf{0.1453} & \textbf{3.068} & 3.832  & 2.68 \\
\bottomrule

\end{tabular}

\caption{\small The comparison between retrieval-based approaches and baseline TTA generation models on the collected 1,525 user prompts (\emph{User}). 
Other approaches include random prompt retrieval from training set using user prompt as queries (\emph{Random}), and using LLM to edit the user prompt directly without exemplars are demonstrations (\emph{LLM}).
The proposed approaches uses in-context editing of prompt with exemplars from (1) \emph{out-domain} text drawn from wiki-103 with fixed length window size of $10$, and from (2) \emph{in-domain} AudioCAPS and BBC sound prompts which AudioLDM learned from. 
We fix the retrieved candidate to up to $K=100$ and experimented with various settings within $K$.
}
\label{tab:main}
\vspace{-2mm}
\end{table*}

\section{Experimental Settings}

We employ AudioLDM~\cite{liu2023audioldm} to generate realistic speech and piano music audio samples. 
We use LLaMa-70B~\cite{wu2023lamini}
as the prompt editing models, which is a decoder-only language model.
We collected and evaluated our approaches $1525$ free-form user prompts (Open-prompts) as real-world test prompts; and evaluated on AudioCAPS~\cite{kim-etal-2019-audiocaps} to see if there is performance degradation if more elaborate, expert-annotated prompts are used instead. 
No training was performed except for the instruction-tuning of the large language models.
Audio samples are evaluated with CLAP~\cite{elizalde2023clap} and FAD~\cite{kilgour2019frechet} for automatic text-audio alignment and distance to clean audios respectively.
Human evaluation was performed in terms of subjective (SUB) and objective (OBJ) human evaluation~\cite{liu2023audioldm,agostinelli2023musiclm} for audio quality assessment by five participants.
Both SUB and OBJ are rated on a scale of 5; and SUB is focused on audio quality and OBJ is measured likewise on a scale of 5 for relevance to the edited prompts, where scores are averaged over the participants.

\begin{table}[h]
\centering
\resizebox{0.7\columnwidth}{!}{%
\begin{tabu}{l | cccc}
\tabucline [1pt]{1-4}
\textbf{Model} & \texttt{\#T} & \texttt{TTR(\%)} & \texttt{$\Delta r_{div}$~$\uparrow$} \\ \hline
Random & \cellcolor{mygray}148  & \cellcolor{mygray}9.56 & 0.474 \\
Farthest  & 147  & 9.61 & 0.422 \\
Closest  & \textbf{117}  & \textbf{10.38} & \textbf{0.602} \\

\tabucline [1pt]{1-4}
\end{tabu}%
}
\caption{\small \#T denote as the number of tokens and TTR(\%) refers to the type token ratios and the prompt divergence ($\Delta r_{div}$).
}
\label{tab:example_analysis}
\end{table}

\vspace*{-3mm}
\section{Results and Analysis}




We first demonstrate that retrieval approach (\emph{exemplar}) synthesize better quality audio samples than the original user prompt baseline (\emph{User}) across automatic metrics and human evaluation in Table~\ref{tab:main}, where consistent improvement is observed across metrics, while $r_{div}$ is reduced with the guidance of demonstration. 
To show that the improvement does not simply come from LLM prompt editing, we also compare \emph{exemplar (K=100, closest)} with \emph{LLM}, where text-audio alignment (CLAP) is increased by $+0.011$ and distance to clean audios (FAD) is further improved by $+2.125$.
Moreover, we revisit the past hypothesis that the most similar exemplars are the best for in-context editing by comparing \emph{exemplar (K=100, farthest)} and exemplar (K=100, random), where exemplars are selected differently from the top-$M$ candidates.
The closest \emph{exemplars} are more distinct examples with highest token-type ratio in Table~\ref{tab:example_analysis}.
Overall, we also found higher agreement of exemplar-based editing as compared with other editing techniques. 
The greatest audio improvement comes from the use of LLM editing with in-context learning, even out-performing pure LLM technique by up to $+0.23$ in subjective icashuman evaluation.



\subsection{Correlation of prompt divergence with audio quality}

\begin{figure}[h]
  \centering
\includegraphics[width=0.8\columnwidth]{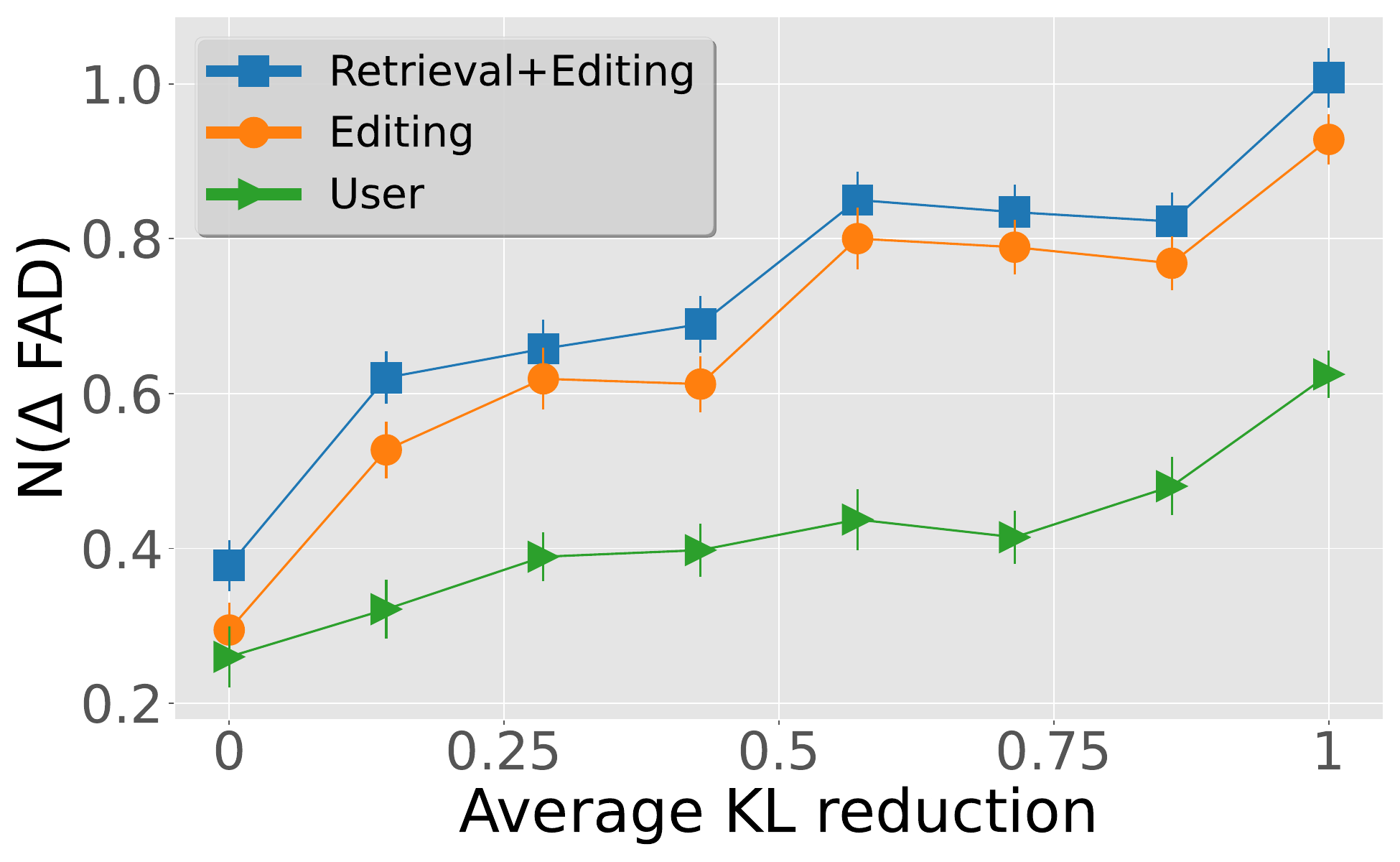}
\caption{ \small 
Plot of average KL reduction on the n-gram feature space, defined as how much the retrieved prompt sets reduces KL divergence to the training distribution.
There is a strong correlation between KL reduction and the audio quality in terms of FAD reduction.
}
\label{fig:dist2}
\end{figure}

Here we intend to show the impact of exemplars from the perspective of prompt divergence metric in Table~\ref{tab:main} and Figure~\ref{fig:dist2}. 
Primarily, the average KL reduction is  linearly proportional to the quality in terms of normalized FAD reduction (N($\Delta$ FAD)), which measures the reduced distance to clean audio samples.
This allows us to deduce the usefulness of retrieved prompts as exemplars in terms of LLM prompt editing.
Further, we compared in- and out-domain prompts in order to show that in-domain prompts as demonstrative exemplars are more effective in driving up the audio qualities. 
We show in Table~\ref{tab:main} this comparison, and see that the domain relevance does indeed help in the editing process, exhibiting a $+0.37$ improvement in FAD and the increases in both human evaluations. 

\subsection{On the inference efficiency}
While in-context retrieval improves audio quality, one major concern remains in terms of its efficiency since the method is employed at inference time. 
In fact, we observe that on a Intel Xeon CPU with FAISS implementation, the average search time for $K=100$ candidate is $2.13$ seconds; and only scale approximately linearly with the total number of training samples.
In general, several factors including
(1) number of clusters: a higher number of cluster corresponds to a better performance, though the difference makes up a minor degree in some instances, as the LLM editing is also crucial in this process. 
(2) The number of retrieved candidates: since we re-compute the similarity of each candidate with user query, the size of the pool will directly (linearly) influence the speed of inference.
(3) Size of dimension: The size of sentence embedding is fixed at $384$ due to S-BERT.

\section{Conclusions}
In this work, we address the challenge of \emph{distributional shift} when the text-to-audio generation models are conditioned on under-specified user prompts.
We propose to edit user prompts with demonstrative exemplars, where training captions are used as demonstrations for the LLMs to better make the edits.
We observed consistent improvement in audio quality as the captions are now closer in distribution to the training captions. 
Our approach is simple and requires no retraining of models, and can be easily adopted to any text-based audio pipelines.

\vfill\pagebreak




\bibliographystyle{IEEEbib}
\bibliography{prompts,music,anthology,custom,acl2021,refs}

\begin{thebibliography}{10}

\bibitem{yang2023diffsound}
Dongchao Yang, Jianwei Yu, Helin Wang, Wen Wang, Chao Weng, Yuexian Zou, and
  Dong Yu,
\newblock ``Diffsound: Discrete diffusion model for text-to-sound generation,''
\newblock {\em IEEE/ACM Transactions on Audio, Speech, and Language
  Processing}, 2023.

\bibitem{liu2023audioldm}
Haohe Liu, Zehua Chen, Yi~Yuan, Xinhao Mei, Xubo Liu, Danilo Mandic, Wenwu
  Wang, and Mark~D Plumbley,
\newblock ``Audioldm: Text-to-audio generation with latent diffusion models,''
\newblock {\em arXiv preprint arXiv:2301.12503}, 2023.

\bibitem{kilgour2019frechet}
Kevin Kilgour, Mauricio Zuluaga, Dominik Roblek, and Matthew Sharifi,
\newblock ``Fr\'echet audio distance: A metric for evaluating music enhancement
  algorithms,'' 2019.

\bibitem{elizalde2023clap}
Benjamin Elizalde, Soham Deshmukh, Mahmoud Al~Ismail, and Huaming Wang,
\newblock ``Clap: Learning audio concepts from natural language supervision,''
\newblock in {\em ICASSP 2023-2023 IEEE International Conference on Acoustics,
  Speech and Signal Processing (ICASSP)}. IEEE, 2023, pp. 1--5.

\bibitem{touvron2023llama}
Hugo Touvron, Thibaut Lavril, Gautier Izacard, Xavier Martinet, Marie-Anne
  Lachaux, Timoth{\'e}e Lacroix, Baptiste Rozi{\`e}re, Naman Goyal, Eric
  Hambro, Faisal Azhar, et~al.,
\newblock ``Llama: Open and efficient foundation language models,''
\newblock {\em arXiv preprint arXiv:2302.13971}, 2023.

\bibitem{min2022rethinking}
Sewon Min, Xinxi Lyu, Ari Holtzman, Mikel Artetxe, Mike Lewis, Hannaneh
  Hajishirzi, and Luke Zettlemoyer,
\newblock ``Rethinking the role of demonstrations: What makes in-context
  learning work?,''
\newblock in {\em Proceedings of the 2022 Conference on Empirical Methods in
  Natural Language Processing}, 2022, pp. 11048--11064.

\bibitem{wei2022chain}
Jason Wei, Xuezhi Wang, Dale Schuurmans, Maarten Bosma, Fei Xia, Ed~Chi, Quoc~V
  Le, Denny Zhou, et~al.,
\newblock ``Chain-of-thought prompting elicits reasoning in large language
  models,''
\newblock {\em Advances in Neural Information Processing Systems}, vol. 35, pp.
  24824--24837, 2022.

\bibitem{DBLP:journals/corr/abs-2005-14165}
Tom~B. Brown, Benjamin Mann, Nick Ryder, Melanie Subbiah, Jared Kaplan,
  Prafulla Dhariwal, Arvind Neelakantan, Pranav Shyam, Girish Sastry, Amanda
  Askell, Sandhini Agarwal, Ariel Herbert{-}Voss, Gretchen Krueger, Tom
  Henighan, Rewon Child, Aditya Ramesh, Daniel~M. Ziegler, Jeffrey Wu, Clemens
  Winter, Christopher Hesse, Mark Chen, Eric Sigler, Mateusz Litwin, Scott
  Gray, Benjamin Chess, Jack Clark, Christopher Berner, Sam McCandlish, Alec
  Radford, Ilya Sutskever, and Dario Amodei,
\newblock ``Language models are few-shot learners,''
\newblock {\em CoRR}, vol. abs/2005.14165, 2020.

\bibitem{liu2019roberta}
Yinhan Liu, Myle Ott, Naman Goyal, Jingfei Du, Mandar Joshi, Danqi Chen, Omer
  Levy, Mike Lewis, Luke Zettlemoyer, and Veselin Stoyanov,
\newblock ``Roberta: A robustly optimized bert pretraining approach,''
\newblock {\em arXiv preprint arXiv:1907.11692}, 2019.

\bibitem{raffel2020exploring}
Colin Raffel, Noam Shazeer, Adam Roberts, Katherine Lee, Sharan Narang, Michael
  Matena, Yanqi Zhou, Wei Li, and Peter~J Liu,
\newblock ``Exploring the limits of transfer learning with a unified
  text-to-text transformer,''
\newblock {\em Journal of Machine Learning Research}, vol. 21, pp. 1--67, 2020.

\bibitem{wu2023large}
Yusong Wu, Ke~Chen, Tianyu Zhang, Yuchen Hui, Taylor Berg-Kirkpatrick, and
  Shlomo Dubnov,
\newblock ``Large-scale contrastive language-audio pretraining with feature
  fusion and keyword-to-caption augmentation,''
\newblock in {\em ICASSP 2023-2023 IEEE International Conference on Acoustics,
  Speech and Signal Processing (ICASSP)}. IEEE, 2023, pp. 1--5.

\bibitem{xie2023data}
Sang~Michael Xie, Shibani Santurkar, Tengyu Ma, and Percy Liang,
\newblock ``Data selection for language models via importance resampling,''
\newblock {\em arXiv preprint arXiv:2302.03169}, 2023.

\bibitem{10.5555/829502.830043}
A.~Broder,
\newblock ``On the resemblance and containment of documents,''
\newblock in {\em Proceedings of the Compression and Complexity of Sequences
  1997}, USA, 1997, SEQUENCES '97, p.~21, IEEE Computer Society.

\bibitem{jaccard1912distribution}
Paul Jaccard,
\newblock ``The distribution of the flora in the alpine zone. 1,''
\newblock {\em New phytologist}, vol. 11, no. 2, pp. 37--50, 1912.

\bibitem{johnson2019billion}
Jeff Johnson, Matthijs Douze, and Herv{\'e} J{\'e}gou,
\newblock ``Billion-scale similarity search with {GPUs},''
\newblock {\em IEEE Transactions on Big Data}, vol. 7, no. 3, pp. 535--547,
  2019.

\bibitem{reimers2019sentencebert}
Nils Reimers and Iryna Gurevych,
\newblock ``Sentence-bert: Sentence embeddings using siamese bert-networks,''
  2019.

\bibitem{kim-etal-2019-audiocaps}
Chris~Dongjoo Kim, Byeongchang Kim, Hyunmin Lee, and Gunhee Kim,
\newblock ``{A}udio{C}aps: Generating captions for audios in the wild,''
\newblock in {\em Proceedings of the 2019 Conference of the North {A}merican
  Chapter of the Association for Computational Linguistics: Human Language
  Technologies, Volume 1 (Long and Short Papers)}, Minneapolis, Minnesota, June
  2019, pp. 119--132, Association for Computational Linguistics.

\bibitem{wu2023lamini}
Minghao Wu, Abdul Waheed, Chiyu Zhang, Muhammad Abdul-Mageed, and Alham~Fikri
  Aji,
\newblock ``Lamini-lm: A diverse herd of distilled models from large-scale
  instructions,''
\newblock {\em arXiv preprint arXiv:2304.14402}, 2023.

\bibitem{agostinelli2023musiclm}
Andrea Agostinelli, Timo~I Denk, Zal{\'a}n Borsos, Jesse Engel, Mauro Verzetti,
  Antoine Caillon, Qingqing Huang, Aren Jansen, Adam Roberts, Marco
  Tagliasacchi, et~al.,
\newblock ``Musiclm: Generating music from text,''
\newblock {\em arXiv preprint arXiv:2301.11325}, 2023.

\end{thebibliography}

\end{document}